\documentclass[preprint2]{aastex}

\usepackage{graphicx}

\slugcomment{Draft - Version 10}

\shorttitle{Anomalous scattering of highly dispersed pulsars}
\shortauthors{L\"ohmer et al.}

\begin{document}


\title{Anomalous scattering of highly dispersed pulsars}


\author{O.~L\"ohmer\altaffilmark{1}, M.~Kramer\altaffilmark{2},
       D.~Mitra\altaffilmark{1}, D.R.~Lorimer\altaffilmark{2},
       A.G.~Lyne\altaffilmark{2}}

\altaffiltext{1}{Max-Planck-Institut f\"ur Radioastronomie,
              Auf dem H\"ugel 69, 53121 Bonn, Germany}
\altaffiltext{2}{University of Manchester, Jodrell Bank Observatory,
              Macclesfield SK11 9DL, UK}



\begin{abstract}
  We report multi--frequency measurements of scatter
  broadening times for nine highly dispersed pulsars over
  a wide frequency range (0.6 -- 4.9~GHz).
  We find the scatter broadening times to be larger than expected
  and to scale with frequency with an average power-law index of $3.44\pm
  0.13$, i.e.~significantly less than that expected from
  standard theories. Such
  possible discrepancies have been predicted very recently by Cordes
  \& Lazio.

\end{abstract}




\section{Introduction}

Fluctuations in the Galactic electron density distribution are
responsible for scintillation and scattering of radio signals
propagating through the interstellar medium (ISM). The resulting
observable phenomena have been studied extensively to investigate the
nature of the density irregularities (e.g.~Rickett
1977\nocite{ric77}).  Pulsars are particularly useful as probes of the
medium because of their small angular diameter and spatial
distribution with samples on many lines of sight through the Galaxy.
Scattering of pulsar signals through an irregular random ISM causes
the signal to arrive from different, multiple ray paths with different
geometric lengths, so that a pulse which left the source at one
instant, arrives at the observer over a typical time interval,
$\tau_{\rm sc}$, commonly the scatter broadening
time. Further, along different ray paths the radiation acquires random
phases which cause interference in the plane of the observer to
produce diffraction patterns. The pattern decorrelates over a
characteristic bandwidth $\Delta \nu_{\rm d}$.

Both the scatter broadening time and decorrelation bandwidth
are strongly dependent on frequency $\nu$, i.e.~$\tau_{\rm sc}
\propto \nu^{-\alpha}$ and $\Delta \nu_{\rm d} \propto \nu^{\alpha}$,
and are related to each other by
$2 \pi \tau_{\rm sc} \Delta \nu_{\rm d} =C$. The
``constant'' $C$ is of the order unity but changes for
different geometries and different models for the turbulence
wavenumber spectrum, $P_{\rm n_e}$ 
(e.g.~Lambert \& Rickett 1999\nocite{lr99}).
Knowledge of the spectrum would provide valuable insight into the
physics of the ISM irregularities. A commonly used description for
the wavenumber spectrum is
a power-law model with large range between ``inner'' and ``outer''
scales, $\kappa_i^{-1}$ and $\kappa_o^{-1}$, (e.g.~Rickett
1977\nocite{ric77}), i.e.
\begin{equation}
\label{eqn:spec}
P_{\rm n_e}(q) = \frac{C^2_{\rm n_e}}{(q^2 +\kappa_o^2)^{\beta/2}} \exp \left[
  -\frac{q^2}{4\kappa_i^2} \right]
\end{equation}
where $q$ is the magnitude of the three-dimensional wavenumber and
$C^2_{\rm n_e}$ is the strength of fluctuation for a given
line-of-sight.  For $\kappa_o\ll q \ll \kappa_i$, one obtains a simple
power-law model with a spectral index $\beta$,
i.e.~$P_{\rm n_e}(q)=C^2_{\rm n_e} q^{-\beta}$ and $\alpha= 2 \beta / (\beta
-2)$ (e.g.~Lee \& Jokipii 1975\nocite{lj75}).  For a pure Kolmogorov
spectrum, $\beta = 11/3$, we expect $\alpha = 4.4$. In turn, by measuring
$\alpha$, one can infer $\beta$ and hence details about the actual
wavenumber spectrum.  A value of $\beta=4$ could, for instance,
describe a medium with abrupt changes in density caused by randomly
placed, discrete clouds along the line-of-sight (Lambert \& Rickett
1999\nocite{lr99}).


\tabcolsep3pt
\begin{deluxetable}{crrcrrrrrrrr}
\tabletypesize{\scriptsize}
\tablecaption{Scatter Broadening for 9 Pulsars \label{tbl1}}
\tablewidth{0pt}
\tablehead{
\colhead{PSR} & \colhead{l} & \colhead{b} &
\colhead{DM} &
\colhead{$\tau_{\rm 0.6}$ } & \colhead{$\tau_{\rm 0.9}$} &
\colhead{$\tau_{\rm 1.4}$ } & \colhead{$\tau_{\rm 1.6}$ } &
\colhead{$\tau_{\rm 2.7}$ } &
\colhead{$\alpha$} & \colhead{$\beta$} & \colhead{$k$} \\
\colhead{} & \colhead{(\degr)} & \colhead{(\degr)} &
\colhead{(pc~cm$^{-3}$)}   &
\colhead{ (ms)} & \colhead{ (ms)} &
\colhead{ (ms)} & \colhead{ (ms)} &
\colhead{(ms)} & & & \\
\colhead{(1)} & \colhead{(2)} & \colhead{(3)} & \colhead{(4)} & \colhead{(5)} &
\colhead{(6)} & \colhead{(7)} & \colhead{(8)} & \colhead{(9)} & \colhead{(10)} &
\colhead{(11)} & \colhead{(12)}
}
\startdata
B1750$-$24 & $4.3 $ & $0.5 $ & $\phn676 $ & & & $49(11)$ & $25(4)$ &
$5.2(1.1)$ & $3.4_{-0.4}^{+0.5}$ & $4.9_{-0.7}^{+1.3}$ & $5.1$ \\
B1758$-$23 & $6.8 $ & $-0.1$ & $1074$ & & & $111(19)$ & $55(10)$ &
$8.6(1.7)$ & $3.9_{-0.4}^{+0.4}$ & $4.1_{-0.4}^{+0.5}$ & $5.7$ \\
B1805$-$20 & $9.5 $ & $-0.4$ & $\phn609 $ & $259(47)$ & $45(9)$ & &
$14(7)$ & & $2.9_{-0.5}^{+0.7}$ & $6.6_{-1.9}^{+4.7}$ & $2.7$\\
B1815$-$14 & $16.4$ & $0.6 $ & $\phn625 $ & & $64(17)$ & $15.0(0.3)$ &
$8.5(0.6)$ & & $3.5_{-0.6}^{+0.5}$ & $4.7_{-0.6}^{+1.4}$ & $2.4$\\
B1817$-$13 & $17.2$ & $0.5 $ & $\phn782 $ & & & $35(3)$ & $19.3(1.7)$ & 
$4.0(1.4)$ & $3.3_{-0.5}^{+0.7}$ & $5.0_{-1.1}^{+1.9}$ & $2.8$\\
B1820$-$11 & $19.8$ & $0.9 $ & $\phn428 $ & $43(6)$ & $11.9(1.9)$ & & & 
& $3.0_{-0.7}^{+0.7}$ & $6.0_{-1.6}^{+9.3}$ & $2.1$\\
B1820$-$14 & $17.3$ & $-0.2$ & $\phn648 $ & $52(51)$ & $10(6)$ &
$2.0(0.4)$ & & & $4.1_{-1.2}^{+0.7}$ & $3.8_{-0.4}^{+1.2}$ & $0.3$\\
B1821$-$11 & $19.8$ & $0.7 $ & $\phn582 $ & $224(180)$ & $40(12)$ &
$8.1(0.8)$ & $3(3)$ & $0.9(1.6)$ & $3.5_{-0.6}^{+0.7}$ &
$4.6_{-0.7}^{+1.4}$ & $2.1$\\
B1849+00 & $33.5$ & $0.0 $ & $\phn680 $ & & & $223(24)$\phd &
$133(11)$ & $36(16)$ & $2.8_{-0.6}^{+1.0}$ & $6.6_{-2.3}^{+7.8}$ & $54.8$\\
\enddata

\tablecomments{Cols.~(2) and (3) give the Galactic longitude
and latitude of each pulsar;
col.~(4) its Dispersion Measure and
cols.~(5)--(9) the scatter broadening with 3$\sigma$-errors in parenthesis
at $0.6, 0.9, 1.4, 1.6$ and $2.7$ GHz. The median of normalized $\chi^2$
for all fits is 1.37. 
Col.~(10) gives the spectral index of scatter broadening
and col.~(11) the spectral index of density irregularities
for each pulsar with their 1$\sigma$--errors.
Col.~(12) gives the ratio $k$ of
$\tau_{\rm sc}$ at 1~GHz scaled by $\alpha$ and 
$\tau_{\rm sc}$ at 1~GHz predicted by
the Taylor \& Cordes (1993)\nocite{tc93} model. Note that, with the
exception of PSR~B1820$-$14, our measured scatter broadening times are
significantly higher than those predicted by the 
Taylor \& Cordes (1993)\nocite{tc93} model.
}
\end{deluxetable}

\begin{figure}
\begin{center}
\begin{tabular}{@{}cc@{}}
\includegraphics[angle=-90,width=6cm]{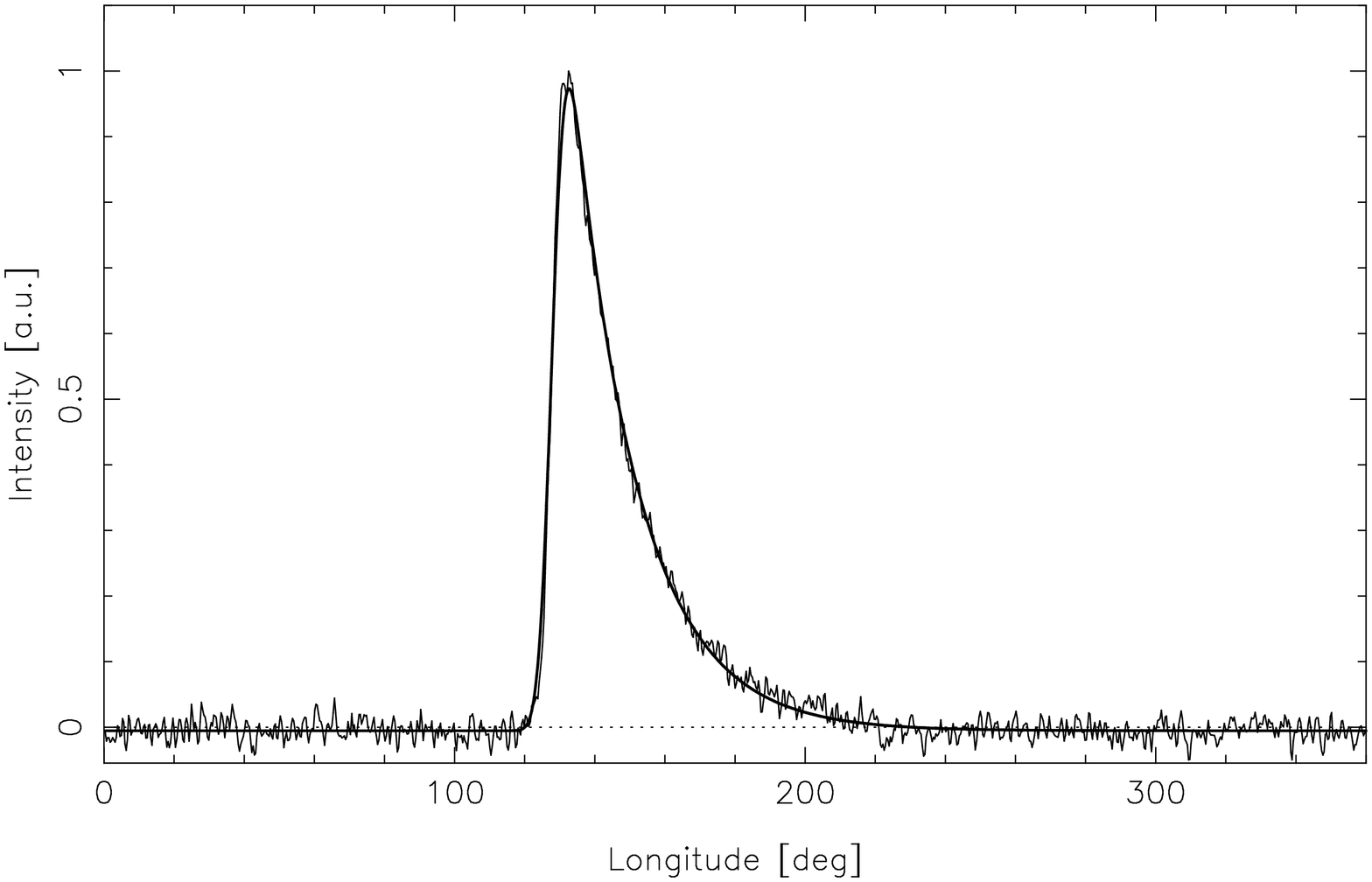}\\
\includegraphics[angle=-90,width=4cm]{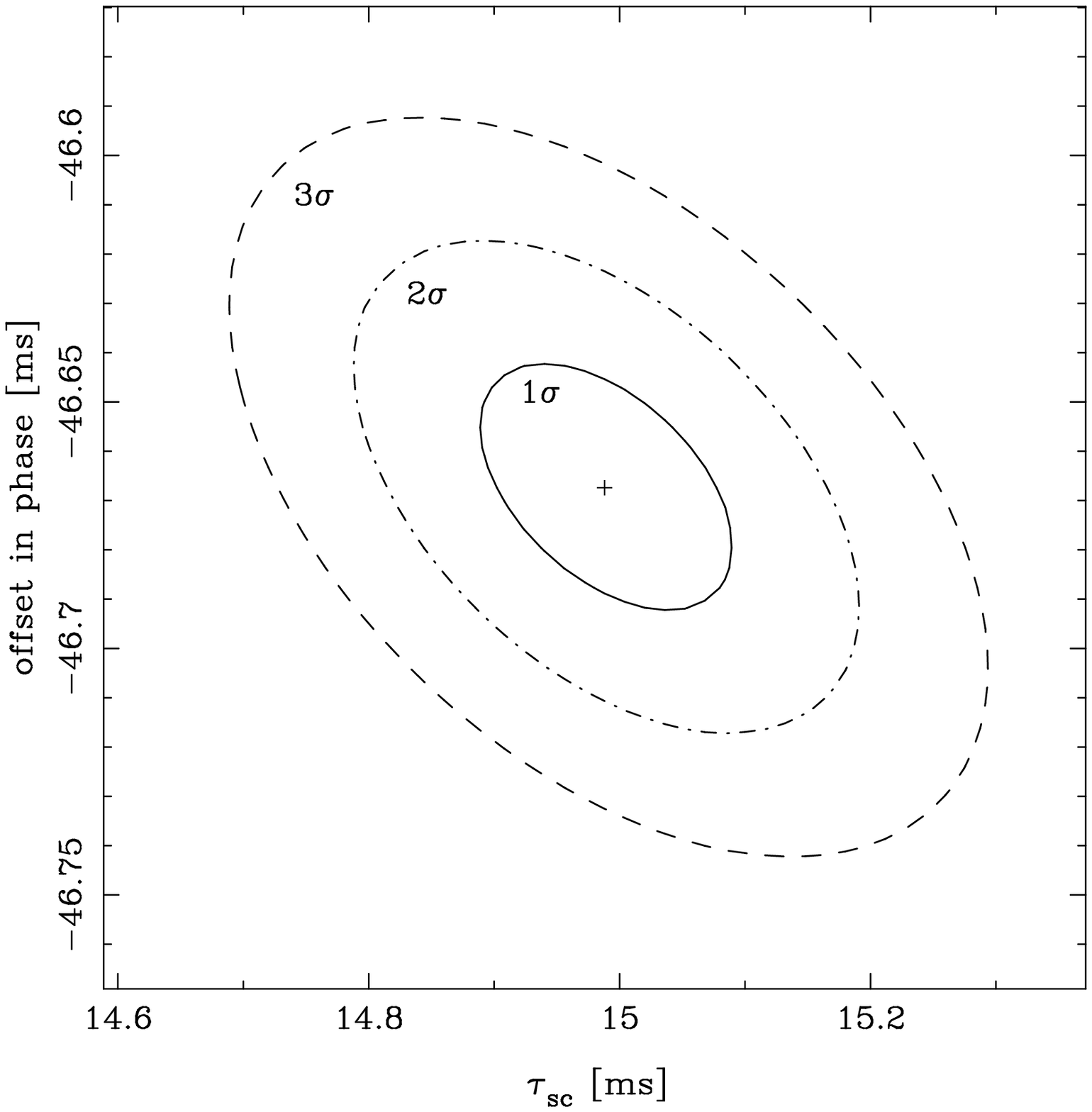}
\end{tabular}
\end{center}
\caption{\label{fig:1815}
({\it Top}) Observed profile of PSR~B1815$-$14 at 1.4~GHz
and best--fit of the model profile (see section~\ref{obs}).
({\it Bottom}) Contours of $\chi^2$ near its global minimum (indicated
by the plus sign) in the
$c$-$\tau_{\rm sc}$ plane for the above profile.
}
\end{figure}


In this {\it Letter}, we report measurements of $\alpha$ by carefully
determining $\tau_{\rm sc}$ over a wide frequency range for a sample
of nine highly dispersed pulsars,
selected to be observable at 4.8~GHz and
located towards the inner
Galaxy. Measurements of $\tau_{\rm sc}$ can be hindered by possible
frequency evolution of the intrinsic pulse shape but we use results from
previous studies of pulsars unaffected by scattering to estimate these pulse
shape effects.  Although one could attempt to measure $\Delta \nu_{\rm
d}$ instead, predicted values are well below 50 Hz at 1 GHz and cannot
be measured with the frequency resolution of typical
filterbanks. This results in a surprisingly small
number of published spectral indices (see Cordes, Weisberg \&
Boriakoff 1985\nocite{cwb85}, Johnston, Nicastro \& Koribalski
1998\nocite{jnk98}). Consequently, the data presented here are in
fact the first ever measurements of $\alpha$ for pulsars
with high dispersion measure (DM).


\section{Observations and Data analysis \label{obs}}

Dual circular polarization observations were made both with the 100-m
radio telescope at Effelsberg (1.4, 2.7 and 4.9 GHz) and the 76-m
Lovell telescope at Jodrell Bank (0.6, 0.9, 1.4 and 1.6 GHz).  Most of
the pulse profiles recorded at Effelsberg were obtained with a
coherently de-dispersing backend, the Effelsberg Berkeley Pulsar
Processor (Backer et al.~1997\nocite{bdz+97}).  For pulsars listed in
Table~\ref{tbl1}, the total available bandwidth ranges from 28 to
45~MHz at 1.4 and 2.7~GHz and up to 128 MHz at 4.85 GHz. At lower
frequencies, we often obtain higher sensitivity by employing an
incoherent hardware de-disperser consisting of $4\times60$ channels of
0.667~MHz, providing a bandwidth of 40~MHz per polarization.
Resulting dispersion broadening times
\footnote{$t_{\rm DM} = 8.3~{\rm DM}~\Delta \nu_{\rm BW} / \nu^3\: \mu s$,
where DM is the dispersion measure (pc cm$^{-3}$),
$\nu$ is the observing frequency (GHz) and $\Delta \nu_{\rm BW}$
the filterbank channel bandwidth (MHz).}, $t_{\rm DM}$,
for the pulsars observed
lie in a range of 0.8~ms $ \le t_{\rm DM} \le $~1.3~ms.
Detailed descriptions of the hardware set-up can be found in
Seiradakis et al.~(1995)\nocite{sgg+95} and Kramer et
al.~(1999)\nocite{kll+99}.

At Jodrell Bank, the pulse profiles were obtained for each
polarization with an incoherent hardware de-disperser using
filterbanks with varying number of channels and bandwidths. At 0.6~GHz
we used 32 channels of 0.125~MHz bandwidth (2.0~ms $ \le t_{\rm DM}
\le $~5.4~ms), at 0.9~GHz 32 channels of 0.25~MHz each (1.1~ms $ \le
t_{\rm DM} \le $~2.0~ms), at 1.4~GHz we employed 32 channels of 1~MHz
each (1.7~ms $ \le t_{\rm DM} \le $~3.2~ms), while we employed 8
channels of 5~MHz bandwidth at 1.6 GHz (0.8~ms $ \le t_{\rm DM} \le
$~10.0~ms). Details of the system can be found in Gould \& Lyne
(1998)\nocite{gl98}.

Typical observation times were between 30 -- 60~min, depending on observing
frequency, bandwidth and flux density of the individual
source. Incoming signals were folded with the topocentric pulse period
and two circular polarizations were added to produce total power
profiles. The signal-to-noise ratios of our profiles range from 13 to
67.

In order to measure the scatter broadening, $\tau_{\rm sc}$, we
performed least--squares fits of an artificial model profile,
$P^{\rm M}(t)$, representing the scattered intrinsic pulse shape, to the
observed profile. For each pulsar, a template $P^{\rm
T}(t)$ is constructed as a sum of Gaussian components fitted to an
unscattered, high--frequency profile using a procedure developed
by Kramer et al. (1994, 1999)\nocite{kwj+94,kll+99}.  Usually, the
high--frequency profile used is observed at 4.9~GHz, except for
PSR~B1820$-$14 and PSR~B1815$-$14 where we use data taken at 1.6 GHz
and 2.7 GHz, respectively, because of the low signal-to-noise of their
4.9~GHz--profiles. For the given DMs, we can expect the
scatter broadening to be negligible at these frequencies (Bhattacharya
et al.~1992\nocite{bwhv92}), but we discuss the validity of this
assumption in more detail later.  For each lower frequency, $P^{\rm
M}(t)$ is the convolution of the template $P^{\rm T}(t)$ with the
impulse response function characterizing the scatter broadening,
$s(t)$, the dispersion smearing across the filterbank channel, $d(t)$,
and the instrumental impulse response, $i(t)$,
\begin{equation}
P^{\rm M}(t) = P^{\rm T}(t) \otimes s(t)\otimes d(t)\otimes i(t)
\end{equation}
where $\otimes$ denotes convolution (Ramachandran, Mitra \& Deshpande
1997\nocite{rmd97}).  The rise times of the receivers and backends are
small enough to consider the effect of $i(t)$ to be negligible, while
$d(t)$ is a rectangular function of width $t_{\rm DM}$ for incoherent
de-dispersion and of zero width for coherent de-dispersion.
Williamson (1972)\nocite{wil72} pointed out that multi--path
scattering due to both thick and thin screens leads to an exponential
decay of the pulse giving $s(t) = {\rm exp}(-t/ \tau_{\rm sc})$.  In
general, $s(t)$ can assume different functional forms for different
geometries reflecting also different rise times of scattered pulses
(Lambert \& Rickett 1999) \nocite{lr99} which could lead to
uncertainties in the estimation of $\tau_{\rm sc}$. However, the
frequency dependence of the scatter broadening time, i.e.~$\alpha$,
can be expected to remain unaffected as we discuss later.


\begin{figure*}
\begin{center}
\includegraphics[angle=-90,width=11cm]{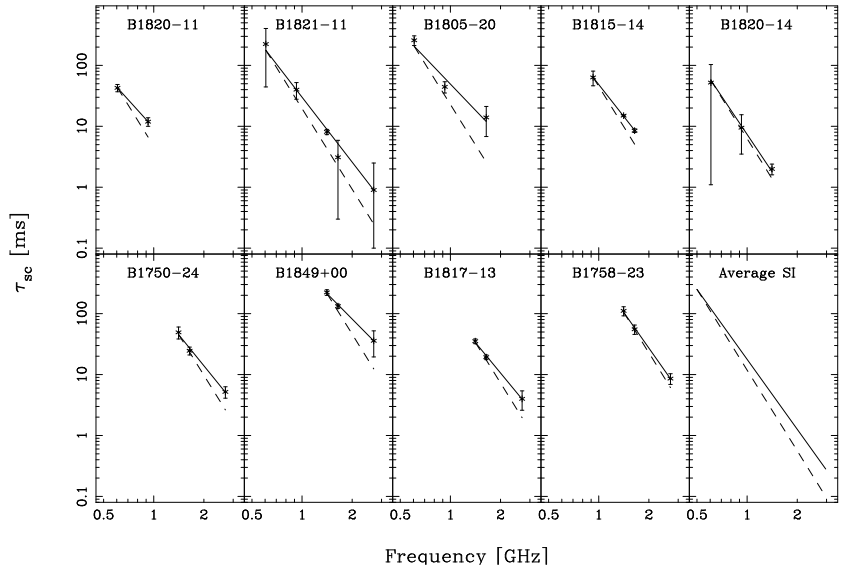}
\end{center}
\caption{ \label{fig:scatfreq}
Scatter broadening times $\tau_{\rm sc}$ with their 3$\sigma$-errors
as a function of observing frequency $\nu$ for nine pulsars.
The lines correspond to the linear fit of the form
$y = \alpha x + K$ where $y = {\rm log}(\tau_{\rm sc})$ in ms and
$x = {\rm log}(\nu)$ in GHz. The dashed lines are examples of the
expected dependence due to a Kolmogorov spectrum, i.e. $\alpha = 4.4$.
The bottom right most panel shows a linear function with an average
spectral index $\langle \alpha \rangle = 3.44$.
}
\end{figure*}

The best fit of the model pulse shape to that observed
is obtained by minimizing the normalized $\chi^2$ value,
\begin{equation}
\chi^2 = \frac{1}{(N-4)\sigma_{\rm off}^2} \sum_{i=1}^{N}
[P^{\rm O}_i -
P^{\rm M}_i(a,b,c,\tau_{\rm sc})]^2 
\end{equation}
where $P^{\rm O}$ is the observed profile, $\sigma_{\rm off}$ the
off-pulse rms, and $N$ the number of bins in the profile. During the
fit, we adjust four free parameters of the model pulse profile $P^{\rm
M}$, i.e.~an amplitude scale-factor, $a$, constant offsets in baseline
\footnote{usually close to zero}, $b$,
and phase, $c$, and the scatter broadening time $\tau_{\rm sc}$.
We obtain best--fit values and uncertainties for $\tau_{\rm sc}$
from the $\chi^2$-contours in the $c$-$\tau_{\rm sc}$ plane.
An example for an observed profile, the best--fit model
profile and $\chi^2$-contours for the fit
is shown in Figure~\ref{fig:1815}.

\begin{figure}
\begin{center}
\includegraphics[angle=-90,width=7cm]{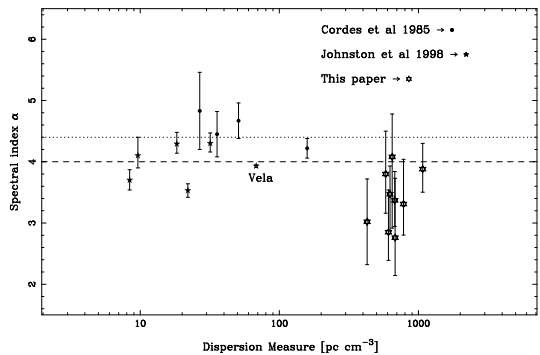}
\end{center}
\caption{ \label{fig:scatdm} Spectral index of scatter broadening
$\alpha$ as a function of DM for our sample of pulsars as well as for
earlier measurements from Cordes et al. (1985) and Johnston et
al.(1998). The dotted line $\alpha = 4.4$ indicates the spectral index
predicted by a pure Kolmogorov spectrum. The dashed line $\alpha =
4.0$ represents the lowest possible index predicted within standard
theories of the turbulent medium.
}
\end{figure}


\section{Factors affecting scatter-broadening-measurements \label{uncert}}

Measuring $\tau_{\rm sc}$ can be difficult since the intrinsic pulse
shape often changes slowly with frequency in two distinct ways. First,
outer profile components often have flatter flux density spectra than
central components (Rankin 1993, Lyne \& Manchester 1988
\nocite{ran83a,lm88}). Second,
pulse widths decrease with increasing frequency (Sieber, Reinecke \&
Wielebinski 1975\nocite{srw75}).
While changes in widths are significant below $\sim$1~GHz, at higher
frequencies the width usually saturates to a constant value
(Thorsett 1991, Xilouris
et al.~1996, Mitra \& Rankin 2001).\nocite{xkj+96,mr01}  
Both effects, if not carefully
accounted for, can give rise to inaccurate estimation of $\tau_{\rm
sc}$. As we use high frequency profiles to create our templates,
stronger outer components could make the template too wide for the
lower frequency profiles.  Due to this effect, we may tend to
underestimate $\tau_{\rm sc}$ at lower frequencies. Alternatively, at
frequencies around 1 GHz, it is also possible to mistake an evolving
trailing profile component for parts of a scattering tail.  This
aspect is however not always present and hard to predict. In contrast,
virtually every pulsar's width is increasing towards
lower frequencies.  The model profile would then be too small,
overestimating $\tau_{\rm sc}$ for frequencies well below 1 GHz.

In order to quantify the impact of these possibly competing effects,
we have simulated ``typical'' pulse profiles and their frequency
evolution, made them subject to scatter broadening and applied our
analysis procedure under the additional presence of noise. While
confirming the above effects, the combined profile evolution hardly
affects our results, i.e.~the derived scatter broadening times agree
very well with the true values. Width evolution does not change the
actual values of the scatter broadening times significantly. The
largest impact is when unnoticed outer components slowly evolve at
high frequencies, which in our sample is only seen for PSRs~B1820$-$11
and B1849+00.  Our simulations show that increased error bars are
sufficient to account for these effects, so that we conservatively
quote 3$\sigma$ error bars for all scatter broadening times in what
follows.  We additionally confirm by determinations of $\alpha$ using
scatter broadening times measured at the two lowest frequencies only
that our results are unaffected by possible changes in pulse profiles.


\section{Results and Discussion \label{results}}

Table~\ref{tbl1} summarizes our measurements of $\tau_{\rm sc}$.  For
PSRs B1750$-$24, B1758$-$23, B1805$-$20, B1815$-$14 and B1820$-$11,
these can be compared to those obtained by Clifton et al.~(1992)
\nocite{clj+92} at a single frequency. We find our more accurate
scatter times to be in very good agreement with their results.  Note
that all but one of the scatter broadening times are factors of
several larger than predicted by the Taylor \& Cordes
(1993)\nocite{tc93} model. However, this model is poorly constrained
for distant, heavily scattered pulsars.

Figure \ref{fig:scatfreq} displays 
the measured frequency dependence of $\tau_{\rm sc}$
from which we 
determine the spectral indices $\alpha$ listed in Table~\ref{tbl1}.
%
%
Earlier multi--frequency observations of pulsars with low DMs (see
Cordes, Weisberg \& Boriakoff 1985\nocite{cwb85}, Johnston, Nicastro
\& Koribalski 1998\nocite{jnk98}) derived spectral indices that are
consistent with a pure Kolmogorov spectrum, i.e.~$\alpha=4.4$
(Figure~\ref{fig:scatdm}).  In contrast, for our highly dispersed
pulsars we find all individual values of $\alpha$ to be
significantly lower than $\alpha=4.4$ and usually even lower than
$\alpha=4.0$ of a ``$\beta=4$--model'' (Lambert \& Rickett 1999). 
The average spectral index is $\langle \alpha \rangle
= 3.44\pm 0.13$ (see also Figure \ref{fig:scatfreq}).

Deriving the spectral index of the wavenumber spectrum from
equation~(\ref{eqn:spec}), $\beta=2\alpha/(\alpha-2)$, we obtain values
that are significantly larger than those expected from standard
theories such as~$\beta=11/3$ or $\beta=4$ (Table~\ref{tbl1}).
Interestingly, Cordes et al.~(1985)\nocite{cwb85} predicted departures
from the Kolmogorov spectrum such as wavenumber enhancements to be
detected in multi-frequency observations of high-DM pulsars. Other
observational inconsistencies with a simple Kolmogorov spectrum have
been explained by steeper spectra ($\beta>4$, e.g.~Goodman \& Narayan
1985\nocite{gn85}), the influence of an inner scale (Coles et
al.~1987\nocite{cfrc87}), or physically slightly different models (Lambert
\& Rickett 1999, 2000\nocite{lr99,lr00}).
It is, however, interesting to note that recent results suggest that
observations along lines of sight of enhanced scattering at low
Galactic latitudes are described by a Kolmogorov spectrum
in an extended medium (e.g.~Stinebring et al.~2000\nocite{ssh+00},
Lambert \& Rickett 2000). While the DMs of the pulsars in these
studies are typically lower, we observe an apparently different
behavior.

Lambert \& Rickett (1999) point out that the scatter broadening
function, $s(t)$, depends to some extent on the actual geometry and
spectral model, although our $\chi^2$ values show that our profiles
can be described accurately by the model.  In the worst case,
according to their results, our individual scatter broadening times
could be consistently under- or over-estimated by an amount which is
typically well covered by our $3\sigma$ error-bars. While $\tau_{\rm
sc}$ computed for a given frequency may change slightly, its spectral
dependence, $\alpha$, shows only minor variations well within the
quoted uncertainties when testing it by increasing (or decreasing)
each $\tau_{\rm sc}$ by a fixed percentage. We conclude that our weak
frequency dependence, $\alpha<4$, is real and cannot be explained by
simple application of standard theories. 

Anisotropic irregularities may lead to variations in the
broadening functions for different frequencies. This causes essentially
the phenomena of {\em anomalous scattering} in the ISM as
discussed recently by Cordes \& Lazio (2001)\nocite{cl01}.  In
particular, one usually assumes that the transverse extent of the
scattering screen is arbitrarily large and that the strength of the
scattering is uniform across the screen. However, if this assumption
is relaxed and/or pulsar distances are large,
several anomalous effects can emerge that can in principle
flatten the observed frequency dependence of $\tau_{\rm sc}$ by
reducing the amount of scattering apparent at {\em lower} frequencies.
In our case, the observed anomalous behaviour could be caused by
scattering at multiple screens with finite extension transverse to
our line-of-sight. As a consequence, less radiation reaches the
observer at lower frequencies since some of the radiation that would
be scattered by an infinite screen is now lost. For such scattering
geometries, one can expect to detect several breaks in the $\tau_{\rm
sc}$ power-law spectra (see e.g.~Figure 3.~of Cordes \& Lazio 2001). A
finer frequency sampling and higher signal-to-noise ratios may be able
to reveal these spectral features in future observations.



\acknowledgments

We are grateful to Jim Cordes, Simon Johnston and Barney Rickett for helpful
discussions.

{\it Note added in proof}. -- The equation $\beta=2\alpha/(\alpha-2)$
is only valid for $\beta<4$.  As shown in Table 1 this results,
however, in values of $\beta$ larger than 4. For turbulence spectra
with $\beta>4$, a relationship $\beta=6-8/\alpha$ is valid (Cordes et
al., 1986, ApJ, 310, 737; Romani et al., 1986, MNRAS, 220,
19). Applying this to $\alpha$ listed in Table 1 results however in
values $\beta<4$. We conclude again that our results are not
consistent with standard theory.  We also note that a scatter
broadening function $s(t) = 1/ \sqrt{t}\; {\rm exp}(-t/ \tau_{\rm
sc})$, which is a conceivable alternative shape in the limit of
one-dimensional scatterers that might pertain to situations with large
scattering (Cordes 2001, private communication), does not appropriately
describe our scattered profiles, in contrast to $s(t)$ used in the
Letter.


\clearpage

\end{document}